\documentclass[sigconf]{acmart}

\AtBeginDocument{%
  }

\copyrightyear{2025}
\acmYear{2025}
\setcopyright{cc}
\setcctype{by}
\acmConference[CCS '25]{Proceedings of the 2025 ACM SIGSAC Conference on Computer and Communications Security}{October 13--17, 2025}{Taipei, Taiwan}
\acmBooktitle{Proceedings of the 2025 ACM SIGSAC Conference on Computer and Communications Security (CCS '25), October 13--17, 2025, Taipei, Taiwan}
\acmDOI{10.1145/3719027.3765079}
\acmISBN{979-8-4007-1525-9/2025/10}

\usepackage{xspace}
\usepackage{subcaption}
\usepackage{fontawesome5}
\usepackage{pifont}
\usepackage{multirow}
\usepackage{makecell}
\usepackage{calc}

\newcommand{\artifactsUrl}{\url{https://doi.org/10.5281/zenodo.17008443}}

\newcommand{\totalKeys}{31,622,338\xspace}
\newcommand{\puttySignatureCount}{58\xspace}

\newcommand{\poneninetwo}{P\nobreakdashes-192\xspace}
\newcommand{\ptwotwofour}{P\nobreakdashes-224\xspace}
\newcommand{\ptwofivesix}{P\nobreakdashes-256\xspace}
\newcommand{\pthreeeightfour}{P\nobreakdashes-384\xspace}
\newcommand{\pfivetwoone}{P\nobreakdashes-521\xspace}

\newcommand{\shaone}{SHA\nobreakdashes-1\xspace}
\newcommand{\shatwo}{SHA\nobreakdashes-2\xspace}
\newcommand{\shatwofivesix}{SHA\nobreakdashes-256\xspace}
\newcommand{\shafiveonetwo}{SHA\nobreakdashes-512\xspace}
\newcommand{\shathree}{SHA\nobreakdashes-3\xspace}

\newcommand{\pkcsone}{PKCS\#1\nobreakdashes~v1.5\xspace}

\newcommand{\puttycve}{CVE-2024-31497\xspace}
\newcommand{\puttyvulnversion}{0.80\xspace}
\newcommand{\puttyfixedversion}{0.81\xspace}
\newcommand{\putty}{PuTTY\xspace}
\newcommand{\nrClients}{24\xspace}

\newcommand{\sshmsg}[1]{\texttt{#1}\xspace}
\newcommand{\msgkexinit}{\sshmsg{KEXINIT}}
\newcommand{\msgnewkeys}{\sshmsg{NEWKEYS}}
\newcommand{\msgkexdhinit}{\sshmsg{KEXDH\_INIT}}
\newcommand{\msgkexdhreply}{\sshmsg{KEXDH\_REPLY}}

\newcommand{\msguserauthrequest}{\sshmsg{USERAUTH\_REQUEST}}

\newcommand{\msguserauthfailure}{\sshmsg{USERAUTH\_FAILURE}}

\newcommand{\supported}{\faCircle}

\newcommand{\unsupported}{\faCircle[regular]}

\newcommand{\fitover}[2]{\makebox[\widthof{#2}][c]{#1}}

\newcommand\faLaunchpad{\raisebox{-0.1\height}{\includegraphics[height=1em]{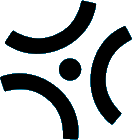}}}

\begin{document}

\title{On the Security of SSH Client Signatures}

\author{Fabian B{\"a}umer}
\orcid{0009-0006-5569-6625}
\affiliation{%
  \institution{Ruhr University Bochum}
  \city{Bochum}
  \country{Germany}
}
\email{fabian.baeumer@rub.de}

\author{Marcus Brinkmann}
\orcid{0000-0001-5649-6357}
\affiliation{%
  \institution{Ruhr University Bochum}
  \city{Bochum}
  \country{Germany}
}
\email{marcus.brinkmann@rub.de}

\author{Maximilian Radoy}
\orcid{0009-0005-3059-6823}
\affiliation{%
  \institution{University Paderborn}
  \city{Paderborn}
  \country{Germany}}
\email{maximilian.radoy@upb.de}

\author{J{\"o}rg Schwenk}
\orcid{0000-0001-9315-7354}
\affiliation{%
  \institution{Ruhr University Bochum}
  \city{Bochum}
  \country{Germany}
}
\email{joerg.schwenk@rub.de}

\author{Juraj Somorovsky}
\orcid{0000-0002-3593-7720}
\affiliation{
  \institution{University Paderborn}
  \city{Paderborn}
  \country{Germany}
}
\email{juraj.somorovsky@upb.de}

\renewcommand{\shortauthors}{Fabian Bäumer, Marcus Brinkmann, Maximilian Radoy, Jörg Schwenk, and Juraj Somorovsky}

\begin{abstract}
  Administrators and developers use SSH client keys and signatures for authentication, for example, to access internet backbone servers or to commit new code on platforms like GitHub. However, unlike \emph{servers}, SSH \emph{clients} cannot be measured through internet scans. We close this gap in two steps. First, we collect \emph{SSH client public keys}. Such keys are regularly published by their owners on open development platforms like GitHub and GitLab. We systematize previous non-academic work by subjecting these keys to various security tests in a longitudinal study. Second, in a series of black-box lab experiments, we analyze the implementations of algorithms for \emph{SSH client signatures} in \nrClients popular SSH clients for Linux, Windows, and macOS.

We extracted \totalKeys keys from three public sources in two scans. Compared to previous work, we see a clear tendency to abandon RSA signatures in favor of EdDSA signatures. Still, in January 2025, we found 98 broken short keys, 139 keys generated from weak randomness, and 149 keys with common or small factors---the large majority of the retrieved keys exposed no weakness.

Weak randomness can not only compromise a secret key through its public key, but also through signatures. It is well-known that a bias in \emph{random nonces} in ECDSA can reveal the secret key through public signatures. For the first time, we show that the use of \emph{deterministic nonces} in ECDSA can also be dangerous: The private signing key of a \putty client can be recovered from just 58~valid signatures if ECDSA with NIST curve \pfivetwoone is used. \putty acknowledged our finding in \puttycve, and they subsequently replaced the nonce generation algorithm.

\end{abstract}

\ccsdesc[500]{General and reference~Measurement}
\ccsdesc[500]{Networks~Transport protocols}
\ccsdesc[500]{Security and privacy~Digital signatures}
\ccsdesc[300]{Security and privacy~Cryptanalysis and other attacks}

\keywords{SSH, Authentication, Digital Signatures, Cryptographic Attacks, Measurement, Public Key Cryptography}

\maketitle

\section{Introduction}
\label{sec:introduction}

System administrators often use the Secure Shell Protocol (SSH) to deploy and manage servers on a private network or the public internet. It also secures other network connections, such as access to Git repositories containing the source code of public or private software projects. Thus, the security of SSH, and especially the security of SSH client authentication, is of critical importance. The IETF standardizes the basic SSH protocol~\cite{rfc4250,rfc4251,rfc4252,rfc4253,rfc4254}, but includes a flexible extension mechanism. Over the years, vendors have added new algorithms, features, and countermeasures to known attacks. So, to understand the real security of SSH, it is not sufficient to analyze standard documents; an investigation of actual implementations and configurations is necessary. While the cryptographic capabilities of clients and servers can be investigated in the lab, the actual \emph{configuration} of SSH clients must be measured in the real world. For servers, this can be done through internet scans (e.g.,~\cite{provos2001scanssh,6838249,USENIX:IzhTeiDur21}).

\paragraph{SSH Clients}

To measure the configuration of SSH clients, we need a strategy that is different from that for servers. Previous work used honeypots~\cite{6624967,melese2016honeypot,Wu2020MiningTI} and observed clients connecting to them via SSH. However, such data is heavily biased toward malicious clients, while we are interested in benign users. We note that SSH clients can authenticate using passwords or digital signatures. While passwords are only sent over an encrypted SSH connection, digital signatures rely on public keys that may be available for analysis. Thus, in our work, we propose to measure SSH clients that use signature-based authentication, which consists of two critical steps: secret key generation and signature generation.

\paragraph{Large-Scale Analysis of SSH Public Keys}

If a secret key is improperly generated, its security can be compromised, given only the public key. For instance, the security of RSA relies on the difficulty of the integer factorization problem. However, if RSA keys are generated incorrectly, different moduli can share a common factor. This opens the door to a batch GCD attack, which can easily factor such vulnerable keys~\cite{bernstein2004find,USENIX:HDWH12}. Similar risks exist with (EC)DSA, where flaws in key generation can lead to severe security vulnerabilities~\cite{cve-2008-0166}. To perform a large-scale analysis of SSH keys, we retrieve SSH public keys published by their owners on GitHub, GitLab, and Launchpad. We then evaluate the distribution of signature algorithms among these keys and test the keys for known weaknesses, such as RSA keys with small factors or a vulnerability to ROCA (Return of Coppersmith's Attack)~\cite{CCS:NSSKM17}. Compared to earlier work analyzing public keys from GitHub and GitLab~\cite{cox2015,EPRINT:Bock23,barbulescu2016weakrsakeys,cryptosense2015,amiet2018} (see \autoref{subsec:longitudinal}), we consider a larger set of keys from more diverse sources, and apply all known, efficient security tests against these data sources at two points in time.

\paragraph{Lab-Evaluation of SSH Client Signatures}

The client key is used in authentication to sign, among other data, the username and a \emph{session identifier} unique to the connection (\autoref{subsec:ssh}). This signature is sent to the server over an encrypted channel and cannot be observed publicly. Thus, we analyze client signatures in the lab. SSH supports RSA, DSA, ECDSA, and EdDSA. The DSA and ECDSA algorithms require a nonce for each signature. Breitner and Heninger~\cite{FC:BreHen19} have shown that weak randomness for nonce generation can lead to a secret key recovery attack, requiring only a small number of signatures for the attacker. To avoid the attack, they recommend the use of deterministic nonces, such as described in RFC~6979~\cite{rfc6979} and recently included in FIPS~186\nobreakdashes-5~\cite{FIPS-186-5}. Using a novel black-box evaluation method, we analyze the (undocumented) methods for (EC)DSA nonce generation in SSH clients. To differentiate between random and deterministic nonces, we have to generate two signatures over the exact same payload, which is impossible in multiple SSH connections due to the use of unique client and server randoms in the session identifier. To overcome this, we let the client connect to a custom server, which triggers a standard-conforming partial authentication success message, provoking the client to authenticate multiple times using the same key over the same connection. This allows us to study the nonce generation algorithm of arbitrary SSH clients and agents over just a network connection (see \autoref{sec:detnoncetest}).

In our evaluation, we found that \putty uses a non-standard deterministic nonce scheme, named \emph{$k_\text{proto}$}. We found that this algorithm outputs highly biased nonces (the nine most significant bits are always zero) when used with the elliptic curve NIST~\pfivetwoone~\cite{nist-sp-800-186}. By applying a solution to the associated hidden number problem, we can compute the private key with a success probability $\geq 0.5$ from just~\puttySignatureCount signatures, and always from~60 signatures.

\paragraph{Dangers of SSH Agents}

\begin{figure*}
    \centering
    \includegraphics[width=\textwidth]{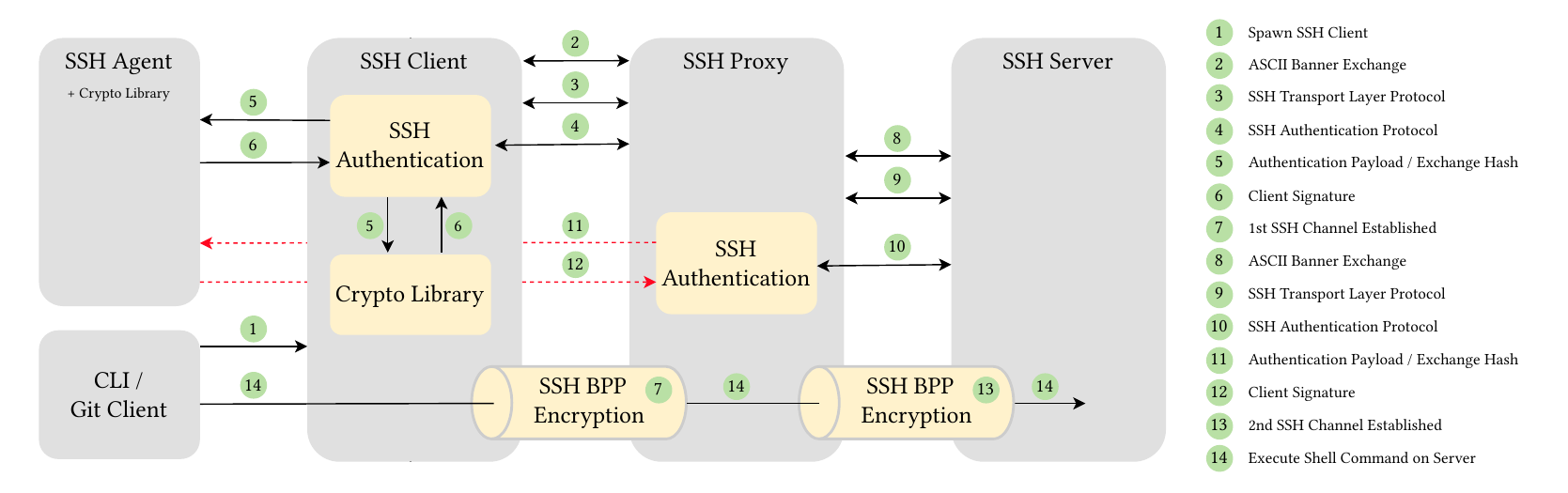}
    \caption{The SSH client ecosystem. (1) The user connects an SSH client to a proxy server, intending to connect to a server only accessible through the proxy. (2-4,7) The SSH protocol runs between the client and the proxy. (5-6) The signature for public key authentication is created either by the client directly or by an SSH agent. (8-10,13) A second SSH protocol runs between the proxy and the intended server. (11-12) The signature is created in the SSH agent local to the user. (14) The user can run shell commands through the two connected channels. As we will show later in \autoref{subsec:sshclientfeatures}, agent forwarding (11-12) is supported by many SSH clients, two of which enable it even by default and provide a signing oracle to SSH proxies.}
    \Description{A figure depicting the complex interactions between different components of the SSH client ecosystem. First, the SSH client is either spawned directly by command invocation or indirectly by, for example, a Git client. The client then connects to the SSH proxy (if any) and performs all steps up to the SSH connection protocol. SSH authentication can be delegated to the SSH agent or cryptographic library as a separate component. Subsequently, the SSH proxy establishes a second connection to the target SSH server. The application data is transmitted through two encrypted BPP channels---from client to proxy and proxy to server.}
    \label{fig:client-eco}
\end{figure*}

Sometimes, SSH connections need to traverse firewalls and overcome network boundaries. In this case, public-facing SSH servers (``jump servers'') can be used as intermediaries, acting as a proxy between the client and the server. There are two options: The proxy server may establish TCP port forwarding to allow for a direct connection between the client and the server. However, port forwarding can also be used to exfiltrate data from an organization's network, and is thus often disabled. Second, the client can use an SSH session to the proxy server and establish a second SSH connection between the proxy and target server from there (\autoref{fig:client-eco}). In this case, the proxy must be able to authenticate itself to the server \emph{as the client}. With signature-based client authentication, this means that the \emph{proxy must be able to sign} the authentication request with \emph{the client's private key}. To enable the proxy to do so, the remote SSH client running on the proxy server can forward the signing request to a local signing oracle, the \emph{SSH agent}. The SSH agent protocol~\cite{ietf-sshm-ssh-agent-01} provides a high-bandwidth signing oracle for arbitrary messages. Thus, users are warned (for example, in the GitHub documentation~\cite{githubdocwarning}) to enable SSH agent forwarding only if they trust the proxy server. However, our analysis revealed two SSH implementations (AbsoluteTelnet and Tectia SSH) that enable SSH agent forwarding by default, ultimately giving all SSH servers that the user connects to the ability to sign arbitrary messages with the user's private key, if not explicitly disabled.

\paragraph{Contributions}

We make the following contributions:
\begin{itemize}
    \item We provide the first longitudinal analysis of SSH client signature keys in the wild, retrieved from Git-based repository hosting services (\autoref{subsec:longitudinal}). 
    We apply a thorough sequence of security tests on the collected keys, revealing 98 broken short keys, 139 keys generated from weak randomness, and 149 keys with common or small factors (\autoref{subsec:security2}).
    \item We evaluate the public key upload restrictions of these hosting services, which can protect users from uploading vulnerable keys to their accounts. We show that all services allow uploading some vulnerable keys, such as short RSA keys, ECDSA keys with points at infinity, or DSA keys with \textit{p} or \textit{q} not prime, among others (\autoref{subsec:upload-restrictions}).
    \item We analyze signature generation algorithms implemented in SSH clients and agents. In particular, we present a novel method to measure an SSH client's (EC)DSA nonce generation algorithm over the network (\autoref{subsec:sshclientfeatures}).
    \item We demonstrate a novel attack on \putty's deterministic nonces for ECDSA,
    recovering a NIST~\pfivetwoone private key from only \puttySignatureCount signatures 
    (\autoref{subsec:putty}).
\end{itemize}

\paragraph{Artifacts}

We share our toolset used in this research paper on Zenodo for permanent access. This toolset includes the SSH client key scraper, evaluation scripts for data aggregation, a script to generate test keys for testing public key upload restrictions, our optimized batch GCD implementation, the source code of the SSH signature sampling and analysis tool, a benchmarking script for the \putty \pfivetwoone private key recovery attack presented in \autoref{subsec:putty}, and additional material. Due to ethical concerns, we do not plan to make the database with all scraped public keys publicly available.
The artifacts are available at:
\begin{quote}
    \artifactsUrl
\end{quote}

\paragraph{Ethical Considerations}
\label{sec:ethics}

We downloaded the public key data from the platforms through their official APIs, adhering to their rate limits. We obtained several factorizations for RSA moduli as a result of our checks on weak keys. To reduce the risk of a potential leak before the platform can revoke the affected keys, we performed the public key collection and analysis on a separate machine with limited access. We did not use these keys in a context other than the described vulnerability analysis; we did not recover or use any private keys to authenticate to any platform.

We disclosed our findings to the platforms, allowing them to inform the affected users, revoke vulnerable keys, and implement upload restrictions as necessary. GitLab responded positively, mitigated the report internally without providing further details, and currently plans to show a warning to users when they upload DSA or weak RSA keys, starting with release 18.3.\footnote{\url{https://gitlab.com/gitlab-org/gitlab/-/issues/432624}, accessed: 2025-07-18} GitHub channels all vulnerability reports to a bug bounty program on HackerOne, whose policy states that users are self-responsible for the content in their repositories, which, to an extent, also applies to SSH public keys. A report that we opened, despite the bug bounties' policy, was closed as out of scope. We conclude that GitHub does not plan to take any further action or refine its upload restrictions. After initial response and disclosure, we did not receive any further response from Canonical regarding public keys on Launchpad.

Although the public keys we analyzed are publicly available, we currently do not plan to publish the complete datasets. The raw datasets contain millions of public keys alongside metadata, including usernames for all existing accounts on each platform and their respective account creation date. This, among other applications, could enable account linking and activity tracking across the three platforms based on the user's public key. While we do provide artifacts that can be used to create similar datasets, we deem the resources necessary to do so significant enough to justify our decision not to publish the raw datasets, thereby thwarting low-effort attacks. Additionally, not publishing the raw datasets maintains the platform operator's ability to restrict access to a user's SSH public keys in the future. The immediate evaluation results contain sensitive information, including partial and complete key factorizations of not yet revoked public keys.

We disclosed our attack on \putty (\puttycve) in early 2024. The \putty maintainer published a fix in \putty \puttyfixedversion. We notified AbsoluteTelnet and Tectia SSH about the default setting for agent forwarding---AbsoluteTelnet intends to change the default behavior in the future.

\section{Background}
\label{sec:background}

\subsection{Attacks on Digital Signature Schemes}
\label{sec:background:sub:dssattacks}

\paragraph{RSA}

A direct way to attack RSA signatures is to compute factors of the RSA modulus, since this would allow attackers to calculate the private key. Notably, when two keys share a prime factor, factoring becomes as easy as calculating the greatest common divisor (GCD) of both moduli; for large collections of RSA moduli, batch GCD algorithms~\cite{bernstein2004find} can be applied. If a weak randomness source is used, this results in a higher probability that a prime factor is reused~\cite{yilek2009private}. RSA is susceptible to \emph{fault attacks} when employing the Chinese Remainder Theorem for modular exponentiation~\cite{JC:BonDeMLip01,CHES:ABFHS02}; such attacks have recently been applied to SSH server authentication~\cite{CCS:RHSH23}. If proprietary optimizations are used in key generation, this may also result in vulnerabilities~\cite{CCS:NSSKM17}. Our complete toolchain to analyze RSA public keys is detailed in \autoref{subsec:securitykeys}.

\paragraph{(EC)DSA}

DSA and ECDSA are parts of the \emph{Digital Signature Standard} (DSS,~\cite{FIPS-186-4,FIPS-186-5}). DSA uses a subgroup $G$ of the multiplicative group of a finite field $GF(p)$ whose order $q$ is tailored to specific hash functions. To compute a DSA signature $(r,s)$, the value $s$ is computed from the hash value $h(m)$ of the message to be signed, the private key $a$ of Alice, and $r=g^k$ computed from a random nonce $k$. It is known that using biased or repeated nonces $k$ enables recovering the private key of Alice from a set of valid DSA signatures~\cite{FC:BreHen19}. To mitigate this threat, the use of deterministic, pseudorandom nonces generated from a high-entropy secret value and some deterministic, non-repeating parameters was proposed~\cite{rfc6979}.

In ECDSA~\cite{FIPS-186-4}, $G$ is an elliptic curve of order $q$. In Appendix D to~\cite{FIPS-186-4}, NIST recommends a list of elliptic curves for ECDSA. While the bit lengths of the elliptic curve moduli \poneninetwo, \ptwotwofour, \ptwofivesix, and \pthreeeightfour align with byte boundaries and hash lengths from the \shatwo and \shathree standards, this is not the case with \pfivetwoone---there is no typo switching the last two digits, but $p=2^{521}-1$ is one of the extremely rare \emph{Mersenne} prime numbers. This special form allows computations in $GF(2^{521}-1)$, and thus in the corresponding elliptic curve group, to be carried out very efficiently. Like their DSA counterparts, ECDSA algorithms allow the private key recovery if the nonces $k$ are biased or reused.

\subsection{SSH}
\label{subsec:ssh}

\paragraph{Standards and Deployment}

IETF standards~\cite{rfc4252,rfc4253,rfc4254} define the basic structure of SSH. There has been no major or minor version update since 2006, despite SSH's constant evolution and the addition of new cryptographic algorithms to its implementations. However, recently, an IETF working group tasked with formally documenting previously undocumented SSH features was established. In contrast to TLS, SSH is mainly deployed in small or closed ecosystems, where a small number of system administrators manage a small number of servers. Large platforms like GitHub offer their key registration services. As a result, no PKI or certificate support is necessary: Both clients and servers store lists of trusted credentials, typically public signature keys or username/password pairs.

\paragraph{SSH Protocol Architecture}

Academic publications typically investigate authenticated key exchange (AKE) protocols, which establish authenticated symmetric keys for client and server, and binary encryption layers, which use these keys for authenticated encryption. The SSH RFCs~\cite{rfc4252,rfc4253,rfc4254} use a different terminology:
\begin{itemize}
    \item The \textbf{Transport Layer Protocol}~\cite{rfc4253} consists of a handshake protocol, where only the server is authenticated, and an authenticated encryption layer called \emph{Binary Packet Protocol (BPP)}.
    During the handshake, the server and client establish session keys for encryption, and the server authenticates itself to the client. To conclude the transport layer protocol, the client typically requests the \emph{authentication protocol} as a subsequent service from the server.
    \item The \textbf{Authentication Protocol}~\cite{rfc4252} offers different methods to authenticate SSH clients and runs within the protected BPP layer of the Transport Layer Protocol. The primary justification for delaying client authentication after the session establishment is to allow the client to authenticate using a password instead of a public key signature. After successful execution, the client has authenticated to the server and specified a service name for the following message exchange. In virtually all cases, this is the connection protocol.
    \item The \textbf{Connection Protocol}~\cite{rfc4254} multiplexes logical channels over the encrypted tunnel. Each channel may be used to implement a single application layer feature, such as remote shells, X11 forwarding, or port forwarding. The protocol runs indefinitely until the connection is closed on either end.
\end{itemize}
In this paper, we focus on the \emph{SSH Authentication Protocol}.
We investigate deployed keys, supported signing algorithms, and the signature creation architecture.

\begin{figure}
    \centering
    \frame{\includegraphics[width=\columnwidth]{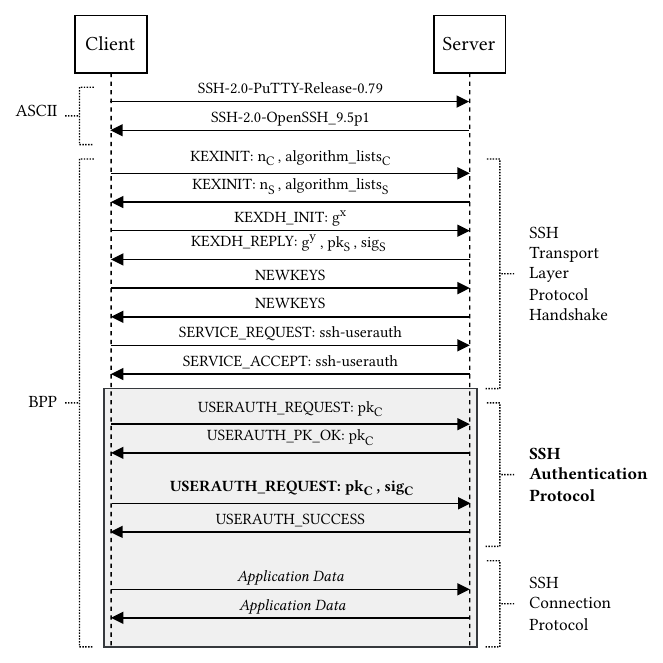}}
    \caption{Typical SSH handshake using a finite-field Diffie-Hellman key exchange. In this paper, we investigate the security of SSH Authentication Protocol implementations, especially the deployed keys and the signature algorithms.}
    \Description{A typical SSH protocol flow using a Diffie-Hellman key exchange between a client and server starting with an ASCII banner exchange, followed by the SSH transport layer protocol handshake consisting out of the algorithm negotiation (\msgkexinit), actual key exchange (\msgkexdhinit / \msgkexdhreply), activation of encryption (\msgnewkeys), and service request by the client. After the handshake, the client authenticates itself as part of the SSH authentication protocol against the server by sending \msguserauthrequest messages with the publickey method. Eventually, application data is transmitted inside the SSH connection protocol.}
    \label{fig:ssh-hs}
\end{figure}

\paragraph{SSH Handshake (\autoref{fig:ssh-hs})}

To initiate an SSH connection, both peers exchange a version banner. The Binary Packet Protocol (BPP) is used from the third message onward, initially without encryption and authentication. In the \msgkexinit messages, nonces and ordered lists of algorithms are exchanged: There is one list for key exchange, one for server signatures, and two (one per direction) each for encryption, MAC, and compression. For each list, the negotiated algorithm is the first algorithm in the client's list, which is also offered by the server. In the \msgkexdhinit and \msgkexdhreply messages, a finite-field Diffie-Hellman key exchange is performed. SSH also supports elliptic curves (ECDH) and hybrid schemes with post-quantum cryptography (PQC) as alternatives. The server authenticates itself with a digital signature as part of the handshake. The signature is computed over the \emph{session identifier}, a hash over the relevant parts of the key exchange.

\subsection{Authentication Protocol Ecosystem}
\label{sec:authentication}

The SSH authentication protocol ecosystem is complex and distributed (\autoref{fig:client-eco}). This section explains its different components, which will be analyzed in \autoref{sec:050-signatures}.

\paragraph{SSH Authentication Protocol}
\label{subsec:SSH-AP}

To authenticate itself after the handshake, the client initiates the SSH Authentication Protocol~\cite{rfc4252} by proposing an authentication method, its username, and other parameters (e.g., password or public key/digital signature). SSH supports a variety of different authentication methods, among these:
\begin{itemize}
    \item \texttt{password} -- The client provides the user's password directly to the server for verification.
    \item \texttt{publickey} -- The client sends a public key, and a signature over the session identifier and authentication request. Supported signature algorithms are based on RSA, DSA, ECDSA, and EdDSA.
    \item \texttt{hostbased} -- Similar to \texttt{publickey}, but the public key is bound to the client's host rather than the user.
    \item \texttt{keyboard-interactive} -- The server may send one or more authentication prompts to the client, which are to be displayed and answered by the user.
\end{itemize}
If the server does not support the proposed authentication method, it may reject and respond with a list of supported methods. The client may request this list directly by proposing the authentication method \verb|none|.

\paragraph{SSH Clients}

\begin{table}
    \centering
    \caption{Data to be signed by the client to perform signature-based client authentication as part of the SSH authentication protocol~\cite[Section 7]{rfc4252}. Data types as per~\cite[Section 5]{rfc4251}.}
    \label{tab:clientsigns}
    \begin{tabular}{ll}
        \toprule
        Data Type        &  Value \\ \midrule
        \texttt{string}  &  session identifier \\
        \texttt{byte}    &  SSH\_MSG\_USERAUTH\_REQUEST\\
        \texttt{string}  &  user name\\
        \texttt{string}  &  service name\\
        \texttt{string}  &  ``publickey''\\
        \texttt{boolean} &  TRUE\\
        \texttt{string}  &  public key algorithm name\\
        \texttt{string}  &  public key to be used for authentication \\
        \bottomrule
    \end{tabular}
\end{table}

If the authentication option \verb|publickey| is chosen, the SSH client authenticates itself via a signature over the exchange hash (similar to the server) plus additional data. In detail, the client signs the data shown in \autoref{tab:clientsigns}. Clients may be configured with multiple public keys using different algorithms to connect to more than one server. Many proprietary implementations of SSH clients exist; for our study, we selected a variety of Linux, Windows, and macOS-based SSH clients (\autoref{tab:clients}).

\paragraph{SSH Servers}

The SSH server is authenticated to the client during the Transport Layer Protocol as part of the handshake. The server's SSH configuration files determine trusted public keys for signature-based client authentication and control which SSH connection protocol features are enabled for each session. For example, SSH access can be limited to a single shell command, or port forwarding can be disabled. The server can advertise the available public key signature algorithms for client authentication in an extension info message~\cite{rfc8308} sent as the first message on the encrypted session before the authentication protocol starts.

\paragraph{SSH Proxies}

Proxies can be used in either of two configurations. As a \emph{jump server}, they allow an authenticated client to access an SSH server on the remote network using port forwarding. In this case, a second SSH connection to a server within an organization's network is tunneled through the first SSH connection to the jump server, which the administrator of the jump server must enable.

Alternatively, the proxy server can act as both an SSH client and server, and thus as a legitimate man-in-the-middle between the SSH client and the destination server. In this case, there are two separate BPP channels: One secure channel between client and proxy (\autoref{fig:client-eco}, 7) and one between proxy and server (\autoref{fig:client-eco}, 13). In the second channel, the proxy (legitimately) can impersonate the client despite public key authentication by making use of \emph{agent forwarding} in the first channel. Agent forwarding provides a signing oracle to the proxy, allowing it to sign the client's authentication request (\autoref{fig:client-eco}, 11) in the second channel without requiring the client to copy its private key files to the proxy.

\paragraph{SSH Agents}

SSH clients may generate digital signatures or delegate signature generation to another program, a so-called \emph{SSH agent}. When used, the SSH agent controls the client's private signing key. The \emph{exchange hash} is computed by the SSH client and then handed over as part of the authentication payload to the SSH agent for signing. Two use cases for SSH agents are mentioned in the literature: (1) to protect the private key of the client by storing it only in the main memory of a separate process, and (2) to enable SSH proxies. SSH agents provide an interface with key management and signing functions~\cite{ietf-sshm-ssh-agent-01}. The primary operations are loading and unloading signing keys, listing the available signing keys, and signing arbitrary data. SSH agents may implement support for constraints on stored keys (for example, requiring explicit user confirmation at the creation of each signature); these constraints require additional configuration. Although agents may limit remote signing to valid user authentication data structures, none of the tested agents enforced any check on the data to be signed---they act as \emph{unconstrained, fully chosen message signing oracles}. SSH agents are popular because they give users access to a passphrase-protected private key without requiring them to enter the passphrase at each login.

\paragraph{Example}

In \autoref{fig:client-eco}, a complex scenario for SSH client authentication is depicted. (1) By some shell/CLI command, an SSH connection to the SSH server is initiated. Since Git can be invoked through a CLI or a GUI that transforms the inputs into CLI commands, our Git use case is covered here. (2) The SSH client will now exchange ASCII banners; since an SSH proxy is configured here, this exchange is between the client and the proxy. (3) The SSH Transport Layer Protocol negotiates algorithms and keys and authenticates the proxy. (4,5,6) In the SSH Authentication Protocol, the client needs a signature over the authentication payload, including the exchange hash. If no proxy is involved, the client has two options: Use its crypto library with its private key or invoke the SSH agent. After establishing the first secure channel (7), the SSH proxy, acting as an SSH client, initiates another banner exchange, this time with the SSH server (8). After the server is authenticated and the BPP parameters are negotiated (9), the SSH proxy can (legitimately) impersonate the client. (11) The authentication payload, including the exchange hash of the second SSH session, is sent to the SSH agent through the first secure BPP channel (7). This is possible because the client has enabled agent forwarding. The SSH proxy receives the signature (12) and completes the protocol (10), establishing a second secure BPP channel (13) between the proxy and server. (14) CLI/shell commands can now be executed on the server securely.

\section{Methodology}
Our methodology is designed to assess the security of SSH client signatures in the wild. As actual client signatures can only be observed by trusted servers, they can not be collected directly. We therefore approach this issue in multiple steps. First, we collect a large number of SSH public keys from Git-based online collaborative development platforms, namely GitHub, GitLab, and Launchpad, as described in \autoref{sec:methodology:sub:collection}. The fact that these platforms reveal the SSH public keys of their users has previously been noted by engineers~\cite{cox2015,golubin2019,filippo2015}, who highlighted the risk to user's privacy.\footnote{As a practical demonstration, Filippo Valsorda maintains a special SSH server \texttt{whoami.filippo.io}, that identifies users trying to connect to it by comparing their public key to a mirror of GitHub's user database.} We then subject these keys to an extensive test suite based on the rules of the CA/Browser forum for TLS server certificates~\cite{ca-browser-forum-2.1.2}, detailed in \autoref{subsec:securitykeys}. Compared to previous work, which limits its scope to specific vulnerabilities or key types, we provide the first comprehensive analysis of all SSH public keys on these platforms. In \autoref{sec:methodology:sub:restrictions}, we analyze these platforms for upload restrictions that are designed as a protection mechanism to prevent users from uploading weak or compromised keys.

Following the analysis of public keys, we also conduct extensive tests of SSH clients and agents in a controlled environment, as described in \autoref{sec:methodology:sub:clients}. In particular, we focus on identifying the underlying cryptographic library and supported algorithms for each analyzed SSH client, followed by checks for deterministic or biased nonces. 

\subsection{Large-Scale Collection of SSH Client Keys}
\label{sec:methodology:sub:collection}

For our analysis in \autoref{sec:clientkeys}, we collect publicly accessible SSH client keys from the online collaborative development platforms GitHub, GitLab, and Launchpad. These platforms commonly use SSH keys to authenticate users accessing source code repositories for reading and writing. We exclude Bitbucket from our study, although it is the third-largest platform, because downloading user SSH keys requires authorized access to workspaces. For each platform, keys are collected through the official GraphQL or REST APIs, respecting the platform's API limits. We use the GraphQL API for GitHub, and a combined approach for GitLab, using GraphQL to list users and REST to retrieve their SSH public keys. Launchpad only offers a REST API.

Each API response is stored in an Elasticsearch database for further processing. As API responses typically list SSH public keys by user, entries are transposed to list users by SSH public key instead. The resulting entries are checked for duplicate keys and, if no duplicate is found, stored in a separate index for analysis. In the case of duplicates, the existing entry is extended, which allows us to analyze public key reuse across multiple user accounts and platforms.

Using this methodology, we retrieve two raw data sets by scraping publicly available SSH keys from GitHub, GitLab, and Launchpad, the first in June 2023 and the second in January 2025 (\autoref{tab:longitudinal}).

\subsection{Security of SSH Client Keys}
\label{subsec:securitykeys}

We analyze each retrieved key according to the rules stated by the CA/Browser forum~\cite[Sections 6.1.1.3, 6.1.5, 6.1.6]{ca-browser-forum-2.1.2}. Although these rules are only intended for TLS server certificates, we believe long-lived SSH client key pairs should also fulfill those requirements related to the security of the key material. We also analyze the set of all RSA keys for common prime factors using the batch GCD attack. To the best of our knowledge, together, these tests cover all known, efficient attacks against public keys that can be run against a large database of keys. Other tests, for example, those found additionally in the RsaCtfTool,\footnote{\url{https://github.com/RsaCtfTool/RsaCtfTool}, accessed: 2025-04-14} are too slow to be run against a large dataset.

\paragraph{Short RSA keys}

\cite[Section 6.1.5]{ca-browser-forum-2.1.2} proposes that the encoded modulus length must be at least 2048 bits and fit byte boundaries. If the modulus $N=pq$ is too small, a brute-force attack that factors $N$ can recover an RSA private key from the public key. State-of-the-art is a number field sieve attack, such as CADO\nobreakdashes-NFS~\cite{cadonfs}. The largest RSA modulus currently factored is RSA\nobreakdashes-250 from the RSA factoring challenge~\cite{rsachallenge} and has 829 bits~\cite{boudot2020}. It was factored by Boudot et al. in 2020, using 2700 CPU-core-years. We consider all keys up to that size broken by brute-force factoring.

\paragraph{RSA Public Exponents}

\cite[Section 6.1.6]{ca-browser-forum-2.1.2} proposes that the public exponent $e$ must be an odd integer larger or equal to $3$ and should be in the range $[2^{16}+1,2^{256}-1]$. A key that is not compliant with this rule is suspect, because a large public exponent $e$ (i.e., roughly the same length as the modulus) may indicate that a small $d$ was chosen during key generation, and $e$ was computed using the extended Euclidean algorithm.
Since small values of $d$ can be found by exhaustive search, we check if the set of public RSA keys contains any large public exponent.

\paragraph{RSA Batch GCD and Small Prime Factors}

In 2012, Heninger et al.~\cite{USENIX:HDWH12} published an attack using a batch GCD algorithm to efficiently find prime factors of individual keys in large sets of RSA keys. The attack exploits cases where RSA keys share a common prime factor. We use an enhanced batch GCD algorithm incorporating subtrees to optimize memory usage, following an approach similar to that proposed by Hastings et al.~\cite{hastings2016}. Our implementation enables the subsequent addition of new keys and the testing of individual keys, with or without their inclusion in the dataset.

According to~\cite[Section 6.1.6]{ca-browser-forum-2.1.2}, the modulus $N$ should not have (prime) factors smaller than $752$. We extend this and check for the first one million prime factors ($p_\text{last}=15{,}485{,}867$) by adding their product during batch GCD computation. For all RSA moduli with one or multiple factors in the first ten thousand primes ($p_\text{last} = 104{,}743$), we apply the elliptic curve factorization algorithm by Lenstra~\cite{lenstra1987factoring} with conservative parameters suitable for finding factors up to 25 decimal digits, as previously done by Cryptosense~\cite{cryptosense2015}.

\paragraph{Debian, ROCA, and Fermat Vulnerabilities}

\cite[Section 6.1.1.3]{ca-browser-forum-2.1.2} states that keys should be checked for known weak keys, such as Debian weak keys~\cite{cve-2008-0166}, ROCA weak keys~\cite{CCS:NSSKM17}, and Fermat weak keys~\cite{EPRINT:Bock23}. In 2008, the OpenSSL version in Debian GNU/Linux had an inadequate procedure to initialize the random number generator. As a result, all RSA and (EC)DSA keys generated with such a system lack entropy and can be brute-forced with only a few hours of CPU time. The badkeys project\footnote{\url{https://badkeys.info/}, accessed: 2025-04-14} by Hanno Böck includes precomputed tables for 1024-bit DSA keys generated by OpenSSH, ECDSA keys under the NIST~\ptwofivesix and \pthreeeightfour curves generated by OpenSSL, as well as RSA keys of size 1024, 2048, 3072, and 4096-bit generated by OpenSSL and OpenSSH. We use code from the badkeys project to check for these keys while simultaneously testing for other known compromised keys. ROCA~\cite{CCS:NSSKM17} describes a type of keys whose prime numbers are generated with reduced entropy. ROCA keys can be detected efficiently with a negligible false positive rate. We use the tool from the original authors to test the keys in our database. The Fermat attack~\cite{EPRINT:Bock23} can find $p$ and $q$ by brute-force search around $\sqrt{N}$ if $p$ and $q$ are close. To achieve this, the attack tries to represent $N = pq$ as the difference of two squares $N = (a + b)(a - b) = a^2 - b^2$ for some $a$ and $b$. To do so, $a$ is initialized to the largest integer $\leq \sqrt{N}$. Then, in what is considered a round in the attack, it is checked whether $a^2 - N$ is a perfect square. If not, $a$ is incremented by $1$, and the attack continues with the next round. We apply the Fermat attack with 100 rounds.

\paragraph{(EC)DSA}

For DSA, we perform an explicit domain parameter validation with partial validation of the generator and a full public key validation as described in~\cite[Section 4.1]{nist-sp-800-89} and~\cite[Section 5.3.1]{nist-sp-800-89}, respectively. Any prime $p$ with fewer than 2048 bits is considered weak. A prime with fewer than 795 bits, the current record of discrete logarithm computation for safe primes~\cite{boudot2019}, is considered broken. Additionally, we check for compromised keys, such as Debian weak keys, using the badkeys tool.

For ECDSA, we first validate that NIST~\ptwofivesix, \pthreeeightfour, or \pfivetwoone is used as the underlying elliptic curve following~\cite[Section 6.1.5]{ca-browser-forum-2.1.2} and that the public key is a valid encoding of an elliptic curve public key. Then, we perform an ECC full public key validation routine described in~\cite[Section 5.6.2.3.3]{nist-sp-800-56a-3} per~\cite[Section 6.1.6]{ca-browser-forum-2.1.2}. This ensures that the public key point is not the identity element, that the coordinates are within valid bounds, that the public key is a point on the curve, and that it has the correct subgroup order. Additionally, we check for compromised keys, such as Debian weak keys, using the badkeys tool.

\paragraph{EdDSA}
For EdDSA public keys, to the best of our knowledge, there are no official security considerations or attacks to consider. Nevertheless, we validate the public key following the decoding process described in~\cite[Section 5.1.3]{rfc8032}, and by calculating the public key point's order. As the decoding process occurs on every signature validation, we expect that keys not passing this test will be unusable in practice. Additionally, we check for compromised keys, such as Debian weak keys, using the badkeys tool.

\subsection{Evaluating Public Key Upload Restrictions}
\label{sec:methodology:sub:restrictions}

To test for existing public key upload restrictions put forth by the platforms, we generate test keys that specifically fail our criteria described in \autoref{subsec:securitykeys}. Each test key should violate a single test criterion to allow for fine-grained detection. Note, however, that this cannot be achieved for all requirements; for example, a key with invalid ECDSA encoding cannot pass any subsequent test. Each test key is then uploaded to the platforms, and the response is recorded. A successful upload is indicated by the platform listing the public key under the user's public keys; a successful authentication is not required.

\subsection{Evaluating SSH Clients in the Lab}
\label{sec:methodology:sub:clients}

We restrict the evaluation of client capabilities to the supported signature algorithms because they determine the strength of the client's authentication.

\paragraph{Selection of Clients and Agents}

We compile a list of SSH clients for evaluation (\autoref{tab:clients}) from a comparison of SSH implementations assembled by Max Horn,\footnote{\url{https://ssh-comparison.quendi.de/}, accessed: 2025-04-14} a Wikipedia page on SSH clients,\footnote{\url{https://wikipedia.org/wiki/Comparison_of_SSH_clients}, accessed: 2025-04-14} and implementations well-known to us. We exclude clients that are no longer actively maintained, as well as clients supporting neither ECDSA nor EdDSA. Evaluation is limited to each client's most recent version, and \nrClients clients in total; evaluating historical versions is out of scope.

For SSH agents, no comparative lists exist. The list in \autoref{tab:agents} is based on internet search results, and includes all agents mentioned on the Wikipedia page on SSH agents.\footnote{\url{https://wikipedia.org/wiki/Ssh-agent}, accessed: 2025-04-14}

\paragraph{Determining Client Features}

For each SSH client implementation, we gather the supported signature algorithms based on the documentation. We also identify the underlying cryptographic library using basic binary analysis, process monitoring, and publicly available information on the vendor's website. For binary analysis under Linux, we use a combination of the tools \texttt{ldd} for identifying required shared libraries, \texttt{nm} for listing static and dynamic symbols, and \texttt{strings} for searching character sequences. Similarly, for binary analysis under Windows, we utilize the \texttt{dumpbin} and \texttt{strings} tools provided by Microsoft. Our findings are detailed in the additional material in the artifacts.

\label{sec:detnoncetest}
\paragraph{Tests for Deterministic/Biased Nonces}

This evaluation is limited to the DSA, ECDSA, and EdDSA algorithms due to the use of a nonce, where even a slight bias in nonce generation can allow for efficient recovery of the secret key. While major SSH implementations have disabled DSA support by default, and the OpenSSH project intends to remove DSA support from its source code entirely in early 2025, some clients still support it for compatibility reasons.

We implement a novel tool in Go that determines the nonce generation method and potential nonce bias, even for proprietary clients without available source code. To do so, the tool captures DSA, ECDSA, or EdDSA client authentication signatures by starting a custom SSH server. If supported by the SSH client, the client is automatically invoked to connect to the server. Otherwise, the connection is set up manually or by using UI automation tools, such as an automatic mouse clicker. We have access to the client's private key, as this tool is designed for in-lab analysis. With the help of this private key, the nonce values are recovered from the sampled signatures and analyzed.

To check for known deterministic nonce schemes from RFC~6979 and \putty \puttyvulnversion ($k_\text{proto}$), our tool generates two nonces according to these schemes and compares them with the nonce value from the received signature. To check for random nonces, the tool enforces further authentication attempts within the same connection, using the same client public key, by indicating a partial authentication success. Partial success is part of the SSH standard and allows for in-protocol, multifactor authentication by sending \msguserauthfailure with a special boolean flag set. Depending on their internal logic, a client may try to authenticate twice or more using the same public key in the same session. Such a client computes the signature twice over the same payload. If these two signatures are identical, we have detected a new, unknown deterministic nonce scheme. Otherwise, we can conclude that the nonce generation is at least partially random.

Our tool can also test for biased nonce generation, applying two statistical tests: a single-bit Z\nobreakdashes-test to detect any single-bit bias, and a more sophisticated Rayleigh test based on an implementation by Daniel Bleichenbacher in Project Wycheproof \footnote{\url{https://github.com/C2SP/wycheproof}, accessed: 2025-04-14} to detect a possible multi-bit bias.
When collecting nonces for this analysis, we do not indicate partial success to avoid introducing bias by repeatedly sampling the same nonce, and instead sample each signature over a fresh payload. The significance level of the test is set to $\alpha = 2^{-32}$ to avoid false positives. For each implementation, a minimum of 10,000 nonces is sampled for bias analysis.

\section{SSH Client Keys}
\label{sec:clientkeys}

\subsection{Longitudinal Analysis of SSH Client Keys}
\label{subsec:longitudinal}

\newcommand{\countstyle}{\small\color{gray}}
\newcommand{\totalstyle}{}
\begin{table*}
    \centering
    \caption{Distribution of asymmetric key algorithms among the SSH client keys on popular source repository platforms.}
    \label{tab:longitudinal}
    \begin{tabular}{lrrrrrrrrr}
        \toprule
        & 2015~\cite{cox2015} & \multicolumn{4}{c}{Our Scan from June 2023} & \multicolumn{4}{c}{Our Scan from January 2025} \\
        \cmidrule(lr){2-2}\cmidrule(lr){3-6}\cmidrule(lr){7-10}
        Algorithm & GitHub & GitHub & GitLab & Launchpad & Unique & GitHub & GitLab & Launchpad & Unique \\
        \midrule
        RSA & 97.67\% & 61.53\% & 79.44\% & 92.22\% & 68.40\% & 50.82\% & 75.79\% & 90.40\% & 60.80\% \\
         & \countstyle{1,205,330} & \countstyle{5,660,641} & \countstyle{4,595,059} & \countstyle{132,215} & \countstyle{9,959,924} & \countstyle{5,338,842} & \countstyle{5,272,676} & \countstyle{136,743} & \countstyle{10,372,614} \\
        Ed25519 & 0.02\% & 37.46\% & 20.08\% & 1.41\% & 30.74\% & 48.24\% & 23.70\% & 3.41\% & 38.39\% \\
         & \countstyle{210} & \countstyle{3,446,461} & \countstyle{1,161,821} & \countstyle{2,026} & \countstyle{4,476,970} & \countstyle{5,066,902} & \countstyle{1,649,163} & \countstyle{5,163} & \countstyle{6,548,889} \\
        ECDSA & 0.07\% & 1.01\% & 0.40\% & 0.60\% & 0.77\% & 0.94\% & 0.44\% & 0.71\% & 0.74\% \\
         & \countstyle{859} & \countstyle{92,903} & \countstyle{23,052} & \countstyle{860} & \countstyle{112,566} & \countstyle{98,350} & \countstyle{30,774} & \countstyle{1,076} & \countstyle{125,593} \\
        DSA & 2.24\% & 0.00\% & 0.08\% & 5.77\% & 0.09\% & 0.00\% & 0.07\% & 5.48\% & 0.07\% \\
         & \countstyle{27,683} & \countstyle{0} & \countstyle{4,757} & \countstyle{8,273} & \countstyle{12,924} & \countstyle{0} & \countstyle{4,676} & \countstyle{8,282} & \countstyle{12,858} \\
        \midrule
        Total & \totalstyle{1,234,082} & \totalstyle{9,200,005} & \totalstyle{5,784,689} & \totalstyle{143,374} & \totalstyle{14,562,384} & \totalstyle{10,504,094} & \totalstyle{6,957,289} & \totalstyle{151,264} & \totalstyle{17,059,954} \\
        \bottomrule
    \end{tabular}
\end{table*}

We collected over 30M SSH client keys from GitHub, GitLab, and Launchpad in two scans in June 2023 and January 2025 (\autoref{tab:longitudinal}). We found that the distribution of algorithms differs considerably among the platforms, with Launchpad users predominantly using RSA keys (92.22\% / 90.40\%) and DSA keys (5.77\% / 5.48\%). In contrast, GitHub and GitLab users also use Ed25519 (37.46\% / 48.24\% and 20.08\% / 23.70\%). Most notably, the percentage and absolute number of Ed25519 keys on GitHub are rising sharply by over 10\% for our datasets, leading to an almost even split between RSA and Ed25519 keys in January 2025. A similar trend can also be observed for GitLab and Launchpad; however, it is less pronounced and far from being split evenly. At the same time, the absolute number of RSA keys on GitHub is receding. 

Compared to earlier reports, we see an increased adoption of elliptic curve algorithms, particularly Ed25519. In 2015, Cartwright-Cox~\cite{cox2015} published data on 1.38M GitHub user keys, of which 1.35M (97.67 \%) are RSA keys, and the remaining 2\% DSA, with only 859 ECDSA and 210 Ed25519 keys. A 2018 study by Romailler et al.~\cite{amiet2018} collected 4.7M GitHub and 1.2M GitLab user keys as part of a much larger dataset of 346M keys. Of that larger dataset, 94.48\% are RSA keys, and less than 1\% are DSA keys. In contrast, we see a strong adoption of Ed25519 paired with a deprecation of DSA on GitHub and GitLab.

\paragraph{Two-Factor Authentication}

Our analysis is based on the cryptographic primitive underlying the SSH key and does not distinguish between regular keys and those stored on U2F devices, as described in~\cite{protocolopensshu2f}. From the SSH algorithm name, we observe sporadic use of Ed25519 and ECDSA NIST~\ptwofivesix U2F keys on GitHub and GitLab only, accounting for approximately 3.5\% of ECDSA NIST~\ptwofivesix keys and less than 1\% of Ed25519 keys.

\paragraph{RSA}

\begin{figure}
    \centering
    \begin{subfigure}{\columnwidth}
        \centering
        \includegraphics[width=\textwidth]{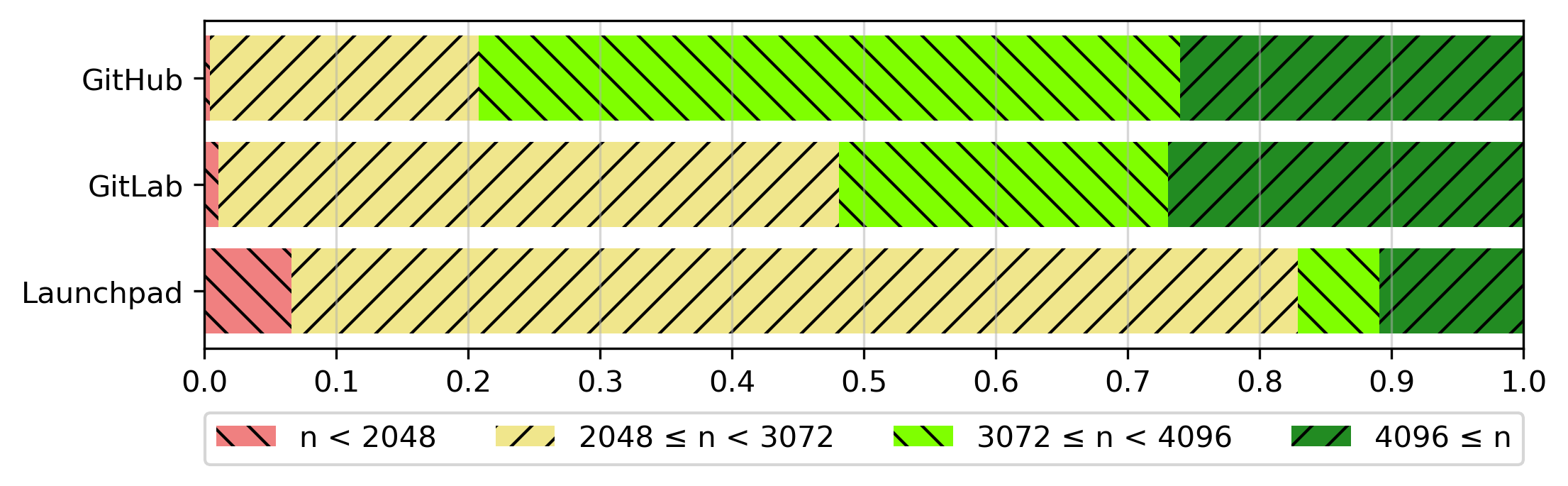}
        \subcaption{Our Scan from June 2023}
    \end{subfigure}
    \begin{subfigure}[b]{\columnwidth}
        \centering
        \includegraphics[width=\textwidth]{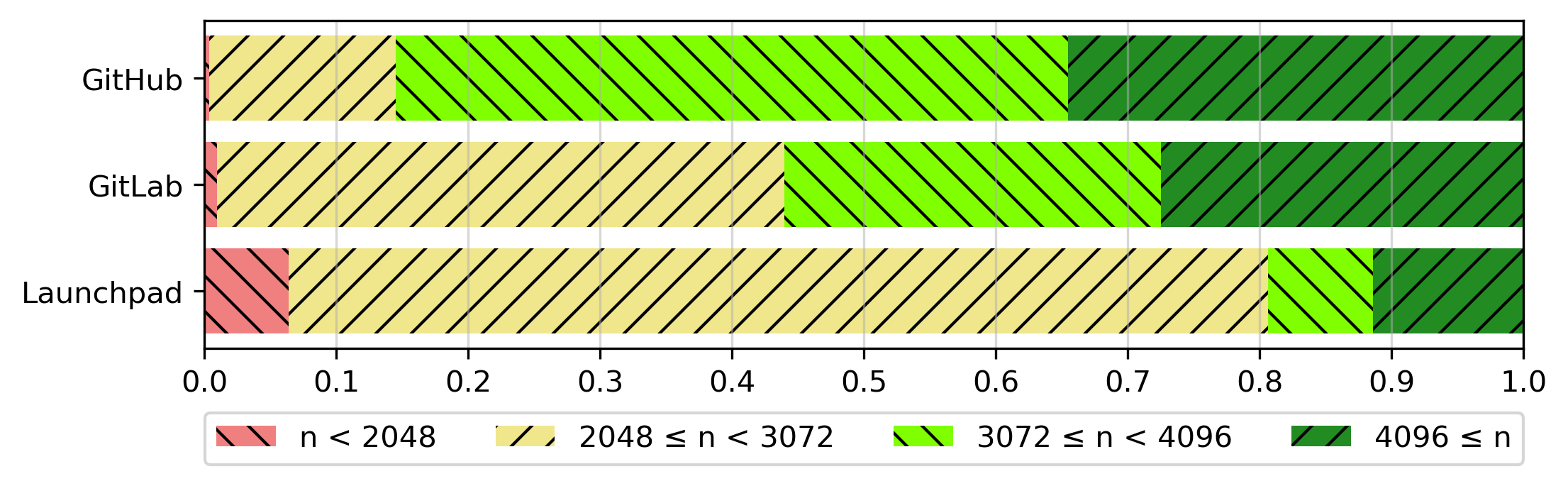}
        \subcaption{Our Scan from January 2025}
    \end{subfigure}
    \caption{The relative distribution of RSA modulus bit lengths $n$ in user keys across each platform shows that GitHub users, by percentage, use the strongest keys. The percentage of longer bit lengths increased measurably between our scans.}
    \Description{Two horizontal bar diagrams comparing the relative distribution of RSA modulus bit lengths found in user keys across all platforms in June 2023 and January 2025. In comparison, GitHub users tend to use longer RSA modulus bit lengths, while a significant portion of users on Launchpad still have keys with n < 2048. Overall, a trend toward more secure bit lengths can be observed, which is more notable on GitHub and GitLab.}
    \label{fig:rsa-mod-dist}
\end{figure}

\autoref{fig:rsa-mod-dist} shows that on GitHub in January 2025, 99.62\% of all RSA keys have a modulus of length at least 2048 bits; this number only amounts to 93.64\% on Launchpad, leaving GitLab in between at 99.08\% in the same time range. A trend towards more secure RSA modulus bit lengths can be observed on all platforms, although it is more noticeable on GitHub and GitLab.

As for the public exponent, $e = 2^{16} + 1$ is the most common choice by a significant margin, accounting for 96.55\% of all exponents in both datasets. Other common choices are $e = 37$ (2.12\%), used by \putty up to version 0.75, and $e = 35$ (1.32\%), used by OpenSSH up to version 5.4. Other values are rarely used for more than one key. The longest RSA public exponent observed has 96 bits alongside a 2158-bit modulus and was found on GitLab.

\paragraph{ECDSA}

\begin{figure}
    \centering
    \begin{subfigure}{\columnwidth}
        \centering
        \includegraphics[width=\textwidth]{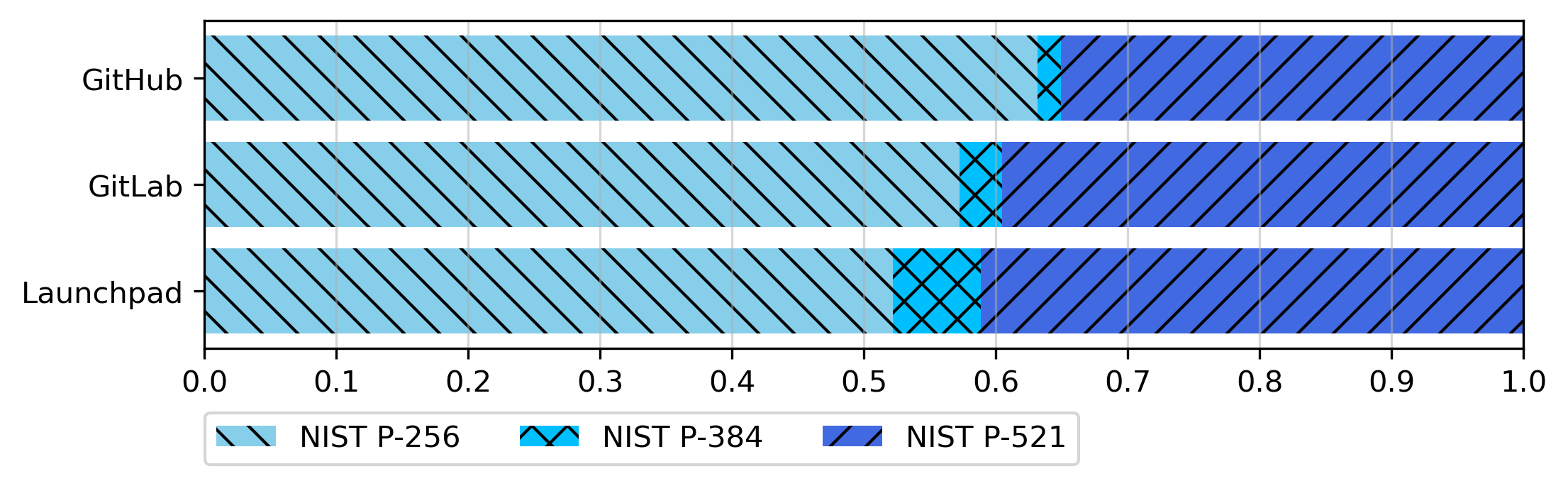}
        \subcaption{Our Scan from June 2023}
    \end{subfigure}
    \begin{subfigure}[b]{\columnwidth}
        \centering
        \includegraphics[width=\columnwidth]{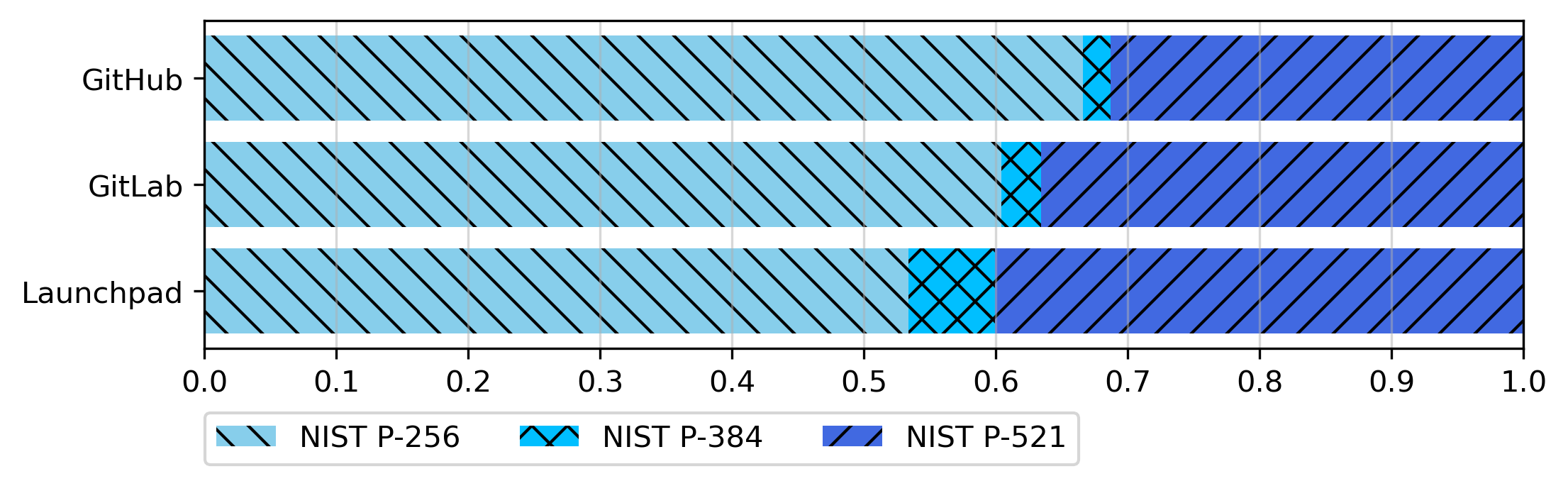}
        \subcaption{Our Scan from January 2025}
    \end{subfigure}
    \caption{The relative distribution of curves used with ECDSA algorithms per platform shows that Launchpad users, by percentage, use more secure ECDSA keys. We observe a shift from NIST~\pfivetwoone towards NIST~\ptwofivesix between our two scans.}
    \label{fig:ecdsa-curve-dist}
    \Description{Two horizontal bar diagrams comparing the relative distribution of curves used with ECDSA algorithms per platform in June 2023 and January 2025. Most users prefer NIST~\ptwofivesix over \pthreeeightfour or \pfivetwoone curves, with the second most common option being \pfivetwoone keys. The amount of \pthreeeightfour keys is significantly less compared to \ptwofivesix and \pfivetwoone. A shift from \pfivetwoone to \ptwofivesix can be observed between the first scan in June 2023 and the second in January 2025.}
\end{figure}

Among all ECDSA keys in our datasets, the most commonly used curves are NIST~\ptwofivesix and NIST~\pfivetwoone, as seen in \autoref{fig:ecdsa-curve-dist}. The only other curve in our dataset is NIST~\pthreeeightfour, with a minor share, thus reflecting the three common choices for ECDSA curves in SSH. Interestingly, we observe a shift from NIST~\pfivetwoone towards less secure NIST~\ptwofivesix keys on all platforms. 

\paragraph{DSA}

We observed only a few thousand DSA public keys available for authentication, none of them on GitHub (cf.~\autoref{subsec:upload-restrictions}). Between our scraping runs, the number of keys on GitLab decreased, while the number on Launchpad stagnated. 88.57\% of all DSA keys use a 1024-bit modulus with a 160-bit subgroup, in line with the specification. However, 10.02\% of keys use moduli of larger bit lengths, most commonly 2048 bits, albeit maintaining the 160-bit subgroup. Using larger moduli sizes is not standard-compliant but is often supported by SSH implementations as long as the subgroup bit length remains unchanged. Due to generic discrete log algorithms (e.g., Pollard's kangaroo~\cite{pollard1978monte} algorithm), increasing the modulus size without increasing the subgroup size does not raise the effective security level.

\subsection{Security of SSH Client Keys}
\label{subsec:security2}

\paragraph{Short RSA keys}

In the June 2023 dataset, we found 77,411 keys with modulus bit lengths shorter than 2048 bits. Of those, 93 keys are shorter than 830 bits, putting them at risk of being factored by computing resources available today. A single key on GitLab was only 8 bits long ($N = 13^2$); probably a test key to probe upload filters on GitLab. As for byte boundary bit lengths, we observe 20,240 keys with a bit length that is not a multiple of 8, potentially causing compatibility issues. About half of these keys are of length $n = 1023$ or $n = 2047$, possibly resulting from a bug in \putty (fixed in version 0.72) that caused some generated RSA keys to be one bit shorter than intended.

We observed similar numbers in January 2025, indicating that the number of short RSA keys has not increased significantly and that users nowadays primarily upload long keys. We found 77,438 keys shorter than 2048 bits, of which 98 are shorter than 830 bits, and 20,692 keys where the modulus bit length is not a multiple of 8. When comparing the two datasets, we found 60,493 shared keys to be shorter than 2048 bits and 90 of those to be shorter than 830 bits. Similarly, 15,777 keys with a bit length not divisible by 8 appear in both datasets.

\paragraph{RSA Public Exponents}

In January 2025, we found four keys with even exponents, three of which have $e = 2^{16} + 2$, and the fourth has a 96-bit exponent, resulting from a corrupt RSA public key; no even exponents were observed in our June 2023 dataset, and thus no such key is shared between the two datasets. In both datasets, all odd exponents had at most 64~bits.\footnote{This is also the limit of public exponent bits for large key sizes in OpenSSL; see \url{https://github.com/openssl/openssl/blob/master/include/openssl/rsa.h\#L56}, accessed: 2025-04-14} We conclude that there is no significant probability for keys with small $d$.

\paragraph{RSA Batch GCD and Small Prime Factors}

In the June 2023 dataset, we identified 305 keys with a GCD greater than one. Among these, 56 were correctly generated keys that reused one of their primes, compromising their security. The remaining 249 keys exhibited small factors, with the largest GCD being 1,842,374,275. Many of the GCDs were smaller, and a substantial portion were composite. Looking only at the first one million prime numbers, we found 306 keys that contained at least one such small prime as a factor. Counting all keys with a GCD greater than one or having a small prime factor, we get a total number of 362 (0.004\%) such keys.

In the January 2025 set, we found 128 keys with a GCD greater than one. Unlike the first set, none of these were correctly generated RSA keys with only two large prime factors. Instead, all keys in this set exhibited small factors, with the largest GCD being 1,300,062,075. Similar to the first set, most GCDs were smaller and predominantly non-prime. Furthermore, 149 keys contained one or more of the first one million prime factors in their modulus. These keys include all 128~keys found with batch GCD, so the total number of keys with a GCD greater than one or having a small prime factor remains 149 (0.001\%).
Of those, 19 keys were found in both datasets.

A potential reason for keys with one or multiple small prime factors can be an accidental change to the Base64 encoding of the public key during upload (e.g., a single Base64 character being altered), rendering those keys unusable \emph{for the legitimate user}. However, we suppose that these keys can still be used for authentication \emph{by an attacker} if the attacker computes a signature using a secret exponent $d$ calculated with the correct value for $\varphi(N)$. From these keys, we selected all keys with a prime factor in the first ten thousand primes, yielding 396 unique keys, 281 keys from June 2023, and 132 keys from January 2025, with 17 keys shared between both datasets. We attempted ECM factorization up to 25-decimal-digit factors on those 396 keys. These attempts returned complete factorizations for 29 keys. Where only partial factorization was successful, the modulus was reduced by 99.53 bits on average.

\paragraph{Debian, ROCA, and Fermat Vulnerabilities}

Our June 2023 dataset contained 93 keys affected by the ROCA vulnerability, compromising their security. All affected keys were found on GitLab or Launchpad. In our January 2025 dataset, this number dropped significantly for GitLab while increasing by one for Launchpad to 25 vulnerable keys across both platforms, 17 of which were already included in our June 2023 dataset. The significant decrease in vulnerable keys on GitLab is likely a response to our first disclosure.

Fermat factorization yielded a single result in our June 2023 dataset, where $N = 13\cdot 13$. Apart from this, the Fermat test with 100 rounds did not return factorizations for any of the other RSA moduli, confirming the negative findings by Böck~\cite{EPRINT:Bock23}.

Additionally, we checked every RSA public key against each \emph{blocklist} defined by the badkeys tool. These blocklists include Debian weak keys for several applications and key sizes, keys vulnerable to CVE-2021-41117, keys leaked in Git repositories and websites, and test keys from RFCs and software tests.
In our June 2023 dataset, we found 176 keys on at least one blocklist. Of these, 73.30\% were blocked as Debian weak keys. Of all keys on Launchpad, 0.08\% are Debian weak keys. This percentage is several orders of magnitude larger than that on the other platforms. The number of keys in a blocklist slightly decreased to 133 RSA keys in our January 2025 dataset, 84.96\% being Debian weak keys. 123 keys are contained in both datasets, of which 106 keys are found on Launchpad, indicating a reduction in blocklisted keys on GitHub and GitLab.

\paragraph{(EC)DSA}

For DSA, we found the same six keys with a bit length of $p$ less than 795 bits on Launchpad in each of the two datasets, which puts them at risk of a discrete logarithm attack with the computing resources available today. Our January 2025 dataset contains one 1024-bit DSA key from Launchpad where $p$ is not prime and $g$ and $y$ do not have order $q$, indicating a corrupt public key. In the same dataset, three unique keys have a subgroup size of 256 bits rather than 160 bits, two of which were already present in our June 2023 dataset. The different subgroup size renders the keys unusable in practice due to the lack of length fields in the DSA signature blob in SSH~\cite[Section 6.6]{rfc4253}. As for Debian weak keys, we found a total of 35 compromised DSA keys across both datasets; none of these keys are shared by different users. However, 29 of these keys appear in both datasets.

For ECDSA, all keys from both datasets passed our validity checks, and the badkeys tool does not find any key on any badkeys blocklist, including Debian weak keys.

\definecolor{RestrictionTableRed}{RGB}{220,0,0}
\definecolor{RestrictionTableGreen}{RGB}{0,150,0}
\newcommand{\RblockedSymbol}{\color{RestrictionTableGreen}\ding{55}}
\newcommand{\Rblocked}{\fitover{\RblockedSymbol}{.}}
\newcommand{\RallowedSymbol}{\color{RestrictionTableRed}\ding{51}}
\newcommand{\Rallowed}{\fitover{\RallowedSymbol}{.}}
\newcommand{\RvalidAllowed}{\fitover{\ding{51}}{.}}
\newcommand{\RvalidBlocked}{\fitover{\ding{55}}{.}}
\begin{table}
    \centering
    \caption{Public key upload restrictions on the Git-based repository hosting services GitHub (\faGithub), GitLab (\faGitlab), and Launchpad (\protect\faLaunchpad) in January 2025. A cross ({\RblockedSymbol}) indicates that the platform prevents the key upload, while a check mark ({\RallowedSymbol}) indicates that the upload is possible. For DSA, $p$ is the prime, $q$ is the order of the subgroup, $g$ is the generator, and $y$ is the public key. For RSA, $N = pq$ is the modulus of bit length $n$, and $e$ is the exponent. For ECDSA, the public key is $(x, y)$.}
    \label{tab:upload-restrictions}
    \small
    \begin{tabular}{p{0em}lcccp{0em}lccc}
        \toprule
         & Test Key & \fitover{\faGithub}{x} & \fitover{\faGitlab}{x} & \fitover{\faLaunchpad}{x} &
         & Test Key & \fitover{\faGithub}{x} & \fitover{\faGitlab}{x} & \fitover{\faLaunchpad}{x}
         \\ \midrule
         \multirow{15}{*}{\rotatebox{90}{RSA}} & Valid & \RvalidAllowed & \RvalidAllowed & \RvalidAllowed & \multirow{10}{*}{\rotatebox{90}{DSA}} & Valid & \RvalidBlocked & \RvalidBlocked & \RvalidAllowed \\
         & $n = 1024$ & \Rallowed & \Rallowed & \Rallowed & & 2048-bit $p$ & \Rblocked & \Rblocked & \Rallowed \\
         & $n = 1023$ & \Rallowed & \Rallowed & \Rallowed & & 256-bit $q$ & \Rblocked & \Rblocked & \Rallowed \\
         & $n = 1022$ & \Rblocked & \Rallowed & \Rallowed & & $p$ not prime & \Rblocked & \Rblocked & \Rallowed \\
         & $n = 512$ & \Rblocked & \Rallowed & \Rallowed & & $q$ not prime & \Rblocked & \Rblocked & \Rallowed \\
         & $8 \nmid n$ & \Rallowed & \Rallowed & \Rallowed & & $g \geq p$ & \Rblocked & \Rblocked & \Rblocked \\
         & Even Modulus & \Rblocked & \Rallowed & \Rallowed & & $y \geq p$ & \Rblocked & \Rblocked & \Rallowed \\
         & Even Exponent & \Rallowed & \Rallowed & \Rblocked & & $ord(g) \neq q$ & \Rblocked & \Rblocked & \Rallowed \\
         & $e < 2^{16} + 1$ & \Rallowed & \Rallowed & \Rallowed & & $ord(y) \neq q$ & \Rblocked & \Rblocked & \Rallowed \\
         & $N = p^2$ & \Rallowed & \Rallowed & \Rallowed & & Debian Weak Key & \Rblocked & \Rblocked & \Rallowed \\
         & Small Factor & \Rblocked & \Rallowed & \Rallowed \\
         & Debian Weak Key & \Rallowed & \Rallowed & \Rallowed \\
         & CVE-2021-41117 & \Rblocked & \Rblocked & \Rallowed & \multirow{7}{*}{\rotatebox{90}{ECDSA}} & Valid & \RvalidAllowed & \RvalidAllowed & \RvalidAllowed \\
         & ROCA & \Rblocked & \Rallowed & \Rallowed & & Invalid Encoding & \Rblocked & \Rblocked & \Rblocked \\ 
         & Fermat & \Rallowed & \Rallowed & \Rallowed & & Point at Infinity & \Rallowed & \Rblocked & \Rblocked \\
         & & & & & & $x \geq p$ & \Rblocked & \Rblocked & \Rblocked  \\
         \multirow{3}{*}{\rotatebox{90}{Ed25519}} & Valid & \RvalidAllowed & \RvalidAllowed & \RvalidAllowed & & $y \geq p$ & \Rblocked & \Rblocked & \Rblocked \\
         & Invalid Encoding & \Rallowed & \Rallowed & \Rallowed & & Not on Curve & \Rblocked & \Rblocked & \Rblocked  \\
         & Point Order $\neq q$ & \Rallowed & \Rallowed & \Rallowed & & Debian Weak Key & \Rallowed & \Rallowed & \Rallowed \\
         \bottomrule
    \end{tabular}
\end{table}

\paragraph{EdDSA}

For EdDSA, we found 327 keys in our June 2023 dataset with an invalid encoding and 315 with a higher-than-expected point order. These numbers increased to 759 invalid encodings and 641 higher point orders in January 2025, where 82 invalid encodings and 80 higher point orders from the June 2023 dataset were also included in the January 2025 dataset. The significant increase in invalid Ed25519 keys is likely caused by the increase in overall Ed25519 keys. Similar to RSA keys, both results may be related to accidental modifications of the Base64-encoded public key during upload. No key with a lower-than-expected point order was observed. Six distinct keys are listed on at least one badkeys blocklist; only one of which is present in both datasets.

\subsection{Public Key Upload Restrictions}
\label{subsec:upload-restrictions}

Our generated test key dataset contains 41 test vectors, including one valid public key per algorithm family. Our results indicate that GitHub does not allow uploading DSA keys, and while some user accounts on GitLab still have DSA keys configured, uploading new DSA keys to GitLab is also no longer possible. For RSA, GitHub prevents uploading short RSA keys with fewer than 1023 bits, keys vulnerable to ROCA, and keys with small factors. Most test vectors for ECDSA were rejected by all platforms, including those with invalid encoding, invalid coordinate bounds, and an invalid point returned invalid key errors. The point at infinity was accepted by GitHub as a public key but rejected by the other two platforms. Only Launchpad returned additional invalid key errors on an out-of-bounds generator for DSA and an even RSA exponent. No upload restrictions were detected for Ed25519 keys on all platforms. The detailed results for all keys can be found in \autoref{tab:upload-restrictions}.

\section{SSH Client Signatures}
\label{sec:050-signatures}

\subsection{SSH Client Features}
\label{subsec:sshclientfeatures}

\autoref{tab:clients} summarizes the basic features of the \nrClients SSH clients we investigated. The most commonly used crypto libraries were OpenSSL~(8), \putty~(4), and LibreSSL~(2)---a fork of OpenSSL. All other libraries appeared only once. All clients support using SSH agents for client authentication; nine are bundled with an agent implementation. All but three libraries support agent forwarding, and two have activated it by default. With regard to nonce generation, we verified that three clients use deterministic nonces for DSA and ECDSA. For three clients, we could not determine whether deterministic nonces are used, as they do not attempt to authenticate multiple times with the same key in the same connection. However, we can rule out either one using $k_\text{proto}$ or RFC~6979 for nonce generation. Regarding biased nonces, we find that the upper nine bits of nonces in \putty~\puttyvulnversion ECDSA NIST~\pfivetwoone signatures are zero, indicating a 9-bit bias. Apart from \putty, we did not find any nonce bias.

For SSH agents (\autoref{tab:agents}), the most commonly used libraries are OpenSSL~(4), \putty~(3), and Libgcrypt~(2). Three use deterministic nonces, and 1Password does not support (EC)DSA. Some Git clients are bundled with SSH clients. This might indicate a slight bias in our dataset towards keys generated by bundled clients.

\begin{table*}
    \centering
    \caption{Overview of SSH clients with their respective versions and operating systems for evaluation. For each client, the table lists whether source code is publicly available (\faCodeBranch), the client ships with an SSH agent implementation (\faUserSecret) or supports the use of agents for authentication (\faSignIn*), and whether agent forwarding is supported (\faShare). A bolt (\faBolt) indicates that agent forwarding is enabled by default, which we consider dangerous. Additionally, for each implementation, the table includes the determined cryptographic library and nonce scheme. For some clients, capturing two signatures for the same plaintext failed due to disconnects after the first authentication attempt. In such cases, we can only rule out known deterministic schemes.}
    \label{tab:clients}
    \begin{tabular}{llcccccll}
         \toprule
         ~ & ~ & ~ & ~ & \multicolumn{3}{c}{SSH Agent} & ~ & ~ \\ \cmidrule{5-7}
         Name & Version & OS & \faCodeBranch & \faUserSecret & \faSignIn* & \faShare & Cryptographic Library & Nonce Scheme \\
         \midrule
         AbsoluteTelnet & 12.16 & \faWindows & \unsupported & \unsupported & \supported & \faBolt & Crypto++ & \textit{Not} RFC 6979 / $k_\text{proto}$ \\
         AsyncSSH & 2.18.0 & \faLinux & \supported & \unsupported & \supported & \supported & OpenSSL 3.3.2 & Random \\
         Bitvise & 9.42 & \faWindows & \unsupported & \unsupported & \supported & \supported & Windows CNG & \textit{Not} RFC 6979 / $k_\text{proto}$ \\
         Cyberduck & 9.0.1 & \faWindows & \supported & \unsupported & \supported & \unsupported & BC Java 1.78.1 & Random \\
         Dropbear & 2024.86 & \faLinux & \supported & \unsupported & \supported & \supported & LibTomCrypt 1.18.2 & Random \\
         Erlang/OTP SSH & 5.2.1 & \faLinux & \supported & \unsupported & \supported & \unsupported & OpenSSL 3.3.2 & \textit{Not} RFC 6979 / $k_\text{proto}$ \\
         FileZilla & 3.67.0 & \faWindows & \supported & \unsupported & \supported & \unsupported & \putty 0.73 & Random \\
         Golang x/crypto/ssh & 0.29.0 & \faWindows & \supported  & \unsupported & \supported & \supported & Go crypto 1.23.3 & Optional (Any) \\
         libssh & 0.11.1 & \faLinux & \supported & \unsupported & \supported & \supported & OpenSSL 3.3.2 & Random \\
         ~ & ~ & ~ & ~ & ~ & ~ & ~ & MBedTLS 3.6.2 & RFC 6979 \\
         OpenSSH Portable & 9.9p1 & \faLinux & \supported & \supported & \supported & \supported & OpenSSL 3.3.2 & Random \\
         ~ & ~ & \faApple & \supported & \supported & \supported & \supported & LibreSSL 3.3.6 & Random \\
         Paramiko & 3.5.0 & \faLinux & \supported & \unsupported & \supported & \supported & OpenSSL 3.3.2 & Random \\
         PKIX-SSH & 15.3 & \faLinux & \supported & \supported & \supported & \supported & OpenSSL 3.3.2 & Random \\
         PortX & 2.2.12 & \faApple & \unsupported & \unsupported & \supported & \supported & OpenSSL 1.1.1 & Random \\
         \putty & \puttyvulnversion & \faWindows & \supported & \supported & \supported & \supported & \putty \puttyvulnversion & $k_\text{proto}$ \\
         ~ & \puttyfixedversion & \faWindows & \supported & \supported & \supported & \supported & \putty \puttyfixedversion & RFC 6979 \\
         SecureCRT & 9.5.2 & \faWindows & \unsupported & \supported & \supported & \supported & BSAFE Crypto-C ME 4.1.5 & Random \\
         Secure ShellFish & 2025.17 & \faApple & \unsupported & \supported & \supported & \supported & OpenSSL & \textit{Not} RFC 6979 / $k_\text{proto}$ \\
         ServerCat & 1.18 & \faApple & \unsupported & \unsupported & \unsupported & \unsupported & BoringSSL & \textit{Not} RFC 6979 / $k_\text{proto}$ \\
         SSH Term & 7.0.26 & \faApple & \unsupported & \unsupported & \unsupported & \unsupported & OpenSSL 3.5.0 & \textit{Not} RFC 6979 / $k_\text{proto}$ \\
         Tectia SSH & 6.6.3.490 & \faWindows & \unsupported & \supported & \supported & \faBolt & \textit{Proprietary} & Random \\
         ~ & ~ & ~ & ~ & ~ & ~ & ~ & OpenSSL 3.0.8 (FIPS)& Random \\
         Tera Term & 5.2 & \faWindows & \supported & \unsupported & \supported & \supported & LibreSSL 3.8.2 & Random \\
         Termius & 9.8.5 & \faLinux & \unsupported & \supported & \supported & \supported & Libsodium 1.0.17 & Random \\
         ~ & 9.21.2 & \faApple & \unsupported & \supported & \supported & \supported & Libsodium 1.0.17 & Random \\
         Win32 OpenSSH & 9.5.0.0 & \faWindows & \supported & \supported & \supported & \supported & LibreSSL 3.8.2 & Random \\
         WinSCP & 6.3.4 & \faWindows & \supported & \unsupported & \supported & \supported & \putty \puttyfixedversion & RFC 6979 \\
         XShell 7 & 0170 & \faWindows & \unsupported & \supported & \supported & \supported & OpenSSL 1.0.2u & Random \\
         \bottomrule
    \end{tabular}
\end{table*}

\begin{table}
    \centering
    \caption{Overview of SSH agent implementations, their underlying cryptographic library, and nonce generation method. 1Password does not support any (EC)DSA algorithms.}
    \label{tab:agents}
    \begin{tabular}{llll}
        \toprule
         Agent & Version & Crypto Library & Nonce  \\ 
         \midrule
         1Password & 8.10.56 & Libgcrypt 1.10.3 & \textit{n/a} \\
         GnuPG & 2.4.4 & Libgcrypt 1.10.3 & RFC 6979 \\
         Goldwarden & 0.3.6 & Go 1.22.1 crypto & Random \\
         KeeAgent & 0.13.8 & BC C\# 1.9.0 & Random \\
         MobaXTerm & 24.4 & \putty \puttyfixedversion & RFC 6979 \\
         OpenSSH & 9.6p1 & OpenSSL 3.0.13 & Random \\
         PKIX-SSH & 15.3 & OpenSSL 3.0.13 & Random \\
         \putty Pageant & \puttyvulnversion & \putty \puttyvulnversion & $k_\text{proto}$ \\ 
         ~ & \puttyfixedversion & \putty \puttyfixedversion & RFC 6979 \\
         SecureCRT & 9.5.2 & \makecell[l]{\scalebox{.67}[1.0]{BSAFE Crypto-C 
         ME 4.1.5}} & Random \\
         Tectia SSH & 6.6.3.490 & \textit{Proprietary} & Random \\
         ~ & ~ & OpenSSL 3.0.8 & Random \\
         Termius & 9.9.0 & Libsodium 1.0.17 & Random \\
         Win32 OpenSSH & 9.5.0.0 & LibreSSL 3.8.2 & Random \\
         XShell 7 Xagent & 0170 & OpenSSL 1.0.2u & Random \\
         \bottomrule
    \end{tabular}
\end{table}

\subsection{\putty P-521 Private Key Recovery}
\label{subsec:putty}

\begin{figure}
	\centering
    \includegraphics[width=\columnwidth]{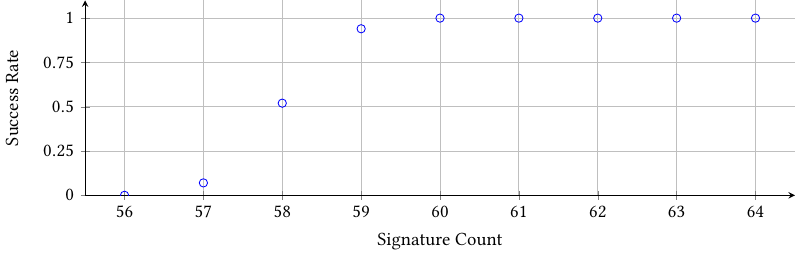}
    \caption[Success rate for the biased nonce attack on \putty ECDSA P-521.]{
      Success rate of the biased nonce attack on \putty ECDSA \pfivetwoone using USVPPredSieve~\cite{EC:AlbHen21}, for a given number of signatures (1024 trials each).
    }
    \Description{A line chart depicting the success rate of biased nonce attack on \putty ECDSA P-521 in relation to the available signature count. At 58 signatures, the success rate is greater than 0.5. At 60 signatures, the success rate is 1.}
    \label{fig:puttyeval}
\end{figure}

\paragraph{\putty's Deterministic Nonce Generation ($k_\text{proto}$)} 

\putty generates nonces deterministically to protect against bad random number generation that could result in biased nonces. The initial implementation in~2001 was designed for DSA, but not ECDSA, and generated the nonce by reducing a 512-bit pseudorandom number $k_\text{proto}$ to a number in the interval $[2,q-1]$. $k_\text{proto}$ is derived from a label, the private key $x$, and the message digest $\text{H}(m)$.\footnote{\url{https://github.com/github/putty/blob/7003b43963aef6cdf2841c2a882a684025f1d806/sshdss.c\#L337}, accessed: 2025-04-14}

\begin{align*}
  \text{digest} &:= \text{SHA-512}\,(\mathit{label} \,||\, x) \\
  k_\text{proto} & := \text{SHA-512}\left(\text{digest} \,||\, \text{H}(m)\right) \\
  k &:= \left(k_\text{proto} \bmod q-2\right) + 2
\end{align*}

For DSA, $q$ is the order of the cyclic subgroup of $\mathbb{Z}_p^*$ generated by the basis element $g$ and is 160 bits in the case of SSH. When support for ECDSA was added in~2014, the code for $k_\text{proto}$ was reused, except for using a different label. In ECDSA, $q$ is the order of the elliptic curve used for all computations generated by the base point $P$.

\paragraph{Bias in P-521}

For DSA, and ECDSA with NIST~\ptwofivesix and \pthreeeightfour, the bias on the nonces $k$ generated with $k_\text{proto}$ is negligible. However, for \pfivetwoone, the output of \shafiveonetwo is always smaller than the size of an element of the underlying field $GF(2^{521}-1)$, and thus the value $k_\text{proto}$ is not reduced. As a result, the nonce $k < 2^{512}$ has the upper nine most significant bits set to zero. 

\paragraph{Hidden Number Problem}

In ECDSA, this nonce $k$ is used in the generation of the signature $(r,s)$ on a message $m$ using the private key $a$ as follows: $r = \text{pr}_x(kP)$ and $s = k^{-1} \cdot (h(m)+a \cdot r) \bmod q$, where $\text{pr}_x$ is the x-coordinate of a point. Rewriting this yields
$$
s^{-1} \cdot r \cdot a + s^{-1}\cdot h(m) \equiv k   \pmod q.
$$
Setting $t = s^{-1} \cdot r$, $b = s^{-1}\cdot h(m)$,  and $\alpha = a$ results in the equivalence relation of the \emph{Hidden Number Problem} (HNP), as formulated by Boneh and Venkatesan~\cite{C:BonVen96},
$$
t \cdot \alpha + b \equiv k \pmod q.
$$
Here, $b$ and $t$ are values known to the attacker, the hidden number $\alpha$ is the secret key the attacker wants to recover, and $k$ is partially known to the attacker, as its nine most significant bits are $0$. Given ``enough'' equivalences, the HNP can be solved, i.e., the value $\alpha = a$ can be computed. The number of HNP equivalences needed to solve the HNP problem (i.e., to compute the private ECDSA signing key) depends on the algorithm used.
 
\paragraph{Lattice Algorithms to Solve the HNP}

For our experiments, we used an algorithm based on lattices from~\cite[USVPPredSieve]{EC:AlbHen21} (see also~\cite{FC:BreHen19}), which can be directly applied to our setting and recovers the private key by solving the Shortest Vector Problem in the constructed lattice. The results are shown in~\autoref{fig:puttyeval}. Given 58 ECDSA signatures generated with $k_\text{proto}$ and \pfivetwoone, we can compute the signer's private key with a probability greater than 0.5. Given 60 such signatures, we can \emph{always} compute the private signing key. We repeated the experiment 1024 times for each signature count, and the computation time for each trial was at most 67 seconds.

\paragraph{Root Cause Analysis}

For the curves \poneninetwo, \ptwotwofour, \ptwofivesix, and \pthreeeightfour, the bit lengths of the coordinates of the curve points were carefully aligned both with byte boundaries and with the output lengths of hash functions: Both hash function families \shatwo and \shathree have hash value lengths of 224, 256, and 384 bits, and the output length 192 appears in NIST SP 800\nobreakdashes-208 as a truncated version of \shatwofivesix. The only exception to this rule is \pfivetwoone, for using a Mersenne prime. Informed programmers wonder about possible typos (512 vs. 521), and discuss why this parameter was chosen.\footnote{\url{https://crypto.stackexchange.com/questions/6219/why-do-the-elliptic-curves-recommended-by-nist-use-521-bits-rather-than-512}, accessed: 2025-04-14}

\section{Related Work}
\paragraph{SSH Client Authentication}

Password-based authentication of SSH clients has been studied previously by Owens and Matthews, and Song et al.~\cite{USENIX:SonWagTia01,owens2008study}. Javed and Paxson~\cite{CCS:JavPax13} statistically measured and evaluated distributed online dictionary attacks against password-based authentication in SSH. To identify client authentication efforts using a stolen SSH client password, Piet et al.~\cite{USENIX:PSPW23} used keystroke authentication. Park et al.~\cite{park2021network} analyzed router log data to detect SSH brute force attacks. SSH honeypots were used by Melese and Avadhani, Koniaris et al., and Wu et al.~\cite{6624967,melese2016honeypot,Wu2020MiningTI} to study brute-force online dictionary attacks. In this paper, we study the security of SSH client authentication with digital signatures.

\paragraph{SSH Client Keys}

In 2015, Cartwright-Cox~\cite{cox2015} scraped 1.2M keys of GitHub users, found 49~RSA keys shorter than 1000~bits, and reported some users whose keys were generated on a Debian system with weak randomness. In 2023, Böck~\cite{EPRINT:Bock23} tested the Fermat attack against this data set and found no vulnerable keys. In 2016, Barbulescu et al.~\cite{barbulescu2016weakrsakeys} analyzed weak RSA keys on the internet, including 3M RSA keys from GitHub users. They found one 512-bit key, but could not find any common factors using batch GCD among the other keys. Also, in 2015, Cryptosense~\cite{cryptosense2015} found small factors in GitHub user keys and could factor~12 keys with 2048~bits using GMP\nobreakdashes-ECM by Zimmermann.\footnote{\url{https://gitlab.inria.fr/zimmerma/ecm}, accessed: 2025-04-14} In 2018, Amiet and Romailler~\cite{amiet2018} analyzed 4.7M keys from GitHub and 1.7M keys from GitLab for ROCA vulnerabilities and common factors using batch GCD. They found vulnerable keys in their dataset, but did not separate results by data source. In this work, we collect keys from GitHub, GitLab, and Launchpad. After reviewing the related work and relevant standards, we identified a broad range of security issues. Based on these issues, we implemented---to the best of our knowledge---the most comprehensive suite of security tests for the detection of potential security issues and vulnerabilities.

\paragraph{SSH Signature-based Authentication}

SSH uses the \pkcsone signature scheme for authentication with RSA keys. Yahyazadeh et al.~\cite{CCS:YCLHDI21} formalized the \pkcsone signature specification and developed a black-box differential testing framework targeting this signature scheme. After analyzing 45 libraries, they identified various conformance issues and six libraries susceptible to Bleichenbacher's signature forgery attacks~\cite{bleichenbacher2006}. These attacks allow an attacker to forge RSA signatures if the verifier does not strictly check the \pkcsone format of the verified signature, and the RSA public key exponent is small (e.g., $e=3$), which we did not observe in the collected data sets. In our work, we concentrated on the security analysis related to signature generation. Signature verification evaluations, such as those presented by Yahyazadeh et al.~\cite{CCS:YCLHDI21}, could extend in our framework.

Ryan et al. recovered private RSA keys of SSH servers from faulty signatures in~\cite{CCS:RHSH23}. Breitner and Heninger studied ECDSA signatures in cryptocurrencies and SSH, and were able to extract private keys due to biased nonces~\cite{FC:BreHen19}. By collecting public SSH server keys from the internet, Heninger et al. could compute the private keys of 0.03\% of SSH servers using RSA and 1.03\% of SSH servers using DSA~\cite{USENIX:HDWH12}. In this paper, we use techniques similar to Heninger et al. to analyze the security of SSH \emph{client} keys.

\paragraph{Deterministic Signature Schemes}

Cao et al. and Poddebniak et al.~\cite{EUROSP:PSSLR18,RSA:CSCCFW22} showed that deterministic variants of ECDSA and EdDSA are vulnerable to \emph{fault attacks}, where invalid signatures are needed. Please note that our attack on \putty is not a fault attack but uses \emph{valid} signatures. We used the algorithm described by Albrecht and Heninger~\cite{EC:AlbHen21} to compute the private key.

\paragraph{Attacks on the SSH Protocol}

In 2009, Albrecht, Paterson, and Watson~\cite{SP:AlbPatWat09} attacked the binary packet protocol of SSH by exploiting the encrypted length field. By measuring the number of ciphertext bytes accepted by the server from the network, they derived a decryption oracle for some bits of an encrypted block. Paterson and Watson~\cite{EC:PatWat10} showed a variation of this attack. Wei Dai~\cite{weidai2002} described a theoretical attack against SSH employing CBC-mode ciphers, while Bhargavan and Leurent~\cite{NDSS:BhaLeu16} presented a handshake transcript collision attack using \shaone. Bäumer et al.~\cite{USENIX:BauBriSch24} showed a prefix truncation attack on the ciphertext for some cipher modes using sequence number manipulation, despite formal security proofs for SSH~\cite{CCS:BelKohNam02,EC:PatWat10,CCS:ADHP16}. Fiterau-Brostean et al.~\cite{fiteruau2017model} learned Mealy state machines of three SSH server implementations; no weaknesses were discovered.

\section{Discussion}
\paragraph{Mitigations}

Users of \putty are advised to update to version \puttyfixedversion or newer to avoid biased nonce generation when using ECDSA signatures with NIST~\pfivetwoone keys. Additionally, users should revoke any NIST~\pfivetwoone keys used with a \putty version below \puttyfixedversion by removing them from servers and platforms and regenerating them as necessary.

Implementers of cryptographic libraries need to carefully avoid biased nonces in (EC)DSA. As our attack on the deterministic nonce generation of \putty has shown, even deterministic nonce schemes may introduce bias. We recommend the algorithm described in~\cite{rfc6979} for the generation of deterministic nonces, which can generate nonces of arbitrary length.

Operators of online collaborative development platforms, and more generally server implementers, are advised to improve their public key security baselines in several aspects. First, they should test public keys on upload and reject keys based on the requirements outlined in~\cite{ca-browser-forum-2.1.2}. To complement this, platforms should perform periodic rescans of all public keys as necessary. While additional upload checks will necessarily add overhead, we estimate that most checks are computationally cheap and therefore reasonable to implement. Platform operators may decide to perform more expensive checks, such as batch GCD, offline.

Furthermore, platforms should reevaluate whether beneficial use cases\footnote{For example, the Ubuntu Linux installer can fetch the user's SSH public key from GitHub using only their GitHub username, providing a convenient way to install their public key on a new system, where copy\&paste may not be available.} for publishing their users' public keys outweigh the negative implications posed by privacy concerns, for example account linking, and scraping for weak keys.

On the SSH client side, vendors need to be aware of the design of the SSH agent interface. Obviously, it can be used by malicious servers to forge SSH signatures, potentially bypassing authentication guarantees. However, because it provides unrestricted access to a perfect signing oracle, it can also be vulnerable to attacks aiming for signature forgeries or private key extraction. Such a signing oracle may allow attackers to sign Git commits impersonating the victim, potentially bypassing repository restrictions. This may lead to unauthorized code or files being added to the repository and could facilitate supply chain attacks. When using modern continuous integration pipelines, this can also enable remote code execution on these systems when commits are built automatically. This can be somewhat mitigated by the agent checking the ``magic bytes'' that are present in Git commits and SSH client signatures. However, since this is a blocklist approach, it may only reduce the attack surface. Also, if side channels exist, for example, through timing, an attacker may use such an oracle to recover the private key used by the SSH agent. To mitigate both potential threats, \texttt{SSH\_AGENT\_CONSTRAIN\_CONFIRM} can be set by default for all keys loaded into an agent, requiring user interaction upon signature generation.

\paragraph{Impact of Weak Keys}

We discovered a variety of weak keys, including RSA keys with fewer than 830 bits and others that were easily factorable---some of which may appear to be merely dummy keys. However, regardless of the specific weakness, a weak key allows an attacker to authenticate itself against the SSH service of a Git-based platform as the affected user. This results in a direct relationship between the impact a weak key can potentially have and the repositories and corresponding privileges a user has access to. If, for example, a user uses the platform solely for private development projects, the impact may be straightforward. If, on the other hand, the user has write access to larger software projects or libraries, this can enable complex supply chain attacks by pushing inconspicuous commits to the repository. We did not explore the potential impact of the weak keys we found in greater detail, because finding all repositories a user has access to is non-trivial. For example, the user might have private repositories not visible to the public.

\section{Conclusions}
In this paper, we shed light on the yet unknown, complex SSH client authentication ecosystem. Among other findings, we show the increasing importance of the EdDSA signature algorithm. We also show the risks associated with proprietary deterministic nonce algorithms in our attack on biased nonces in the ECDSA NIST~\pfivetwoone implementation of \putty.

We have also shown that weak or corrupt SSH public keys can pose a significant threat to the integrity of a user's account. In a longitudinal scan of three source repository platforms, we have found keys compromised by various attacks. We have also shown that upload filters scanning for vulnerable keys can significantly reduce the user's exposure to such vulnerabilities. Platforms that deploy such filters have a substantially lower number of affected weak keys among their users.

\begin{acks}
We thank Kira Höltgen, whose master's thesis included an early version of the key scraper as well as the collection and responsible disclosure of the June 2023 dataset. Fabian Bäumer was supported by the German Federal Ministry of Research, Technology and Space (BMFTR) project ``Combinatorial testing of TLS libraries at all levels (KoTeBi)'' (16KIS1557). Marcus Brinkmann was supported by the Deutsche Forschungsgemeinschaft (DFG, German Research Foundation) under Germany's Excellence Strategy - EXC 2092 CASA - 390781972.

\end{acks}

\bibliographystyle{ACM-Reference-Format}
\balance
\bibliography{bib/abbrev2,bib/crypto,bib/rfc,bib/paper}

\end{document}